\documentclass[aps,prl,twocolumn,showpacs,longbibliography]{revtex4-1}
\usepackage{color}
\usepackage{amssymb}
\usepackage{amsmath}
\usepackage{bm}
\usepackage{graphicx}

\begin{document}
\newcommand{\ndp}{n_{p}}
\newcommand{\ddp}{d_{p}}
\newcommand{\mdp}{m_{p}}
\newcommand{\Vdp}{V_{p}}
\newcommand{\Idp}{I_{p}}
\newcommand{\rhop}{\rho_{p}}
\newcommand{\rhof}{\rho_{f}}
\newcommand{\Gal}{\textit{Ga}}
\newcommand{\Atw}{\textit{A}}
\newcommand{\Sto}{\textit{St}}
\newcommand{\Rey}{Re}
\newcommand{\Ret}{\Rey_t}
\newcommand{\Rep}{\Rey_p}
\newcommand{\mlbot}{z_0}
\newcommand{\mltop}{z_1}
\newcommand{\mlcent}{c}
\newcommand{\bbot}{\mltop}
\newcommand{\btop}{z_2}
\newcommand{\bcent}{c_b}
\newcommand{\PhiV}{\Phi}
\newcommand{\Phitop}{\phi_{b}}
\newcommand{\taugrav}{\tau}
\newcommand{\taug}{\tau_g}
\newcommand{\taut}{\tau_t}
\newcommand{\vg}{v_g}
\newcommand{\vt}{v_t}
\newcommand{\mlwid}{h}
\newcommand{\bwid}{h_b}
\newcommand{\Zavg}{{xy}}
\newcommand{\ftop}{{fp}}
\newcommand{\ptof}{{pf}}
\newcommand{\ptop}{{pp}}
\newcommand{\gridsp}{\Delta x}
\title{A new form of mixing in turbulent sedimentation}

\author{Simone Tandurella$^{1}$}

\author{Marco Edoardo Rosti$^{1}$}
\email{marco.rosti@oist.jp}

\author{Stefano Musacchio$^{2}$}

\author{Guido Boffetta$^{2}$}

\affiliation{$^{1}$Complex Fluids and Flows Unit, Okinawa Institute of Science and Technology Graduate University, 1919-1 Tancha, Onna-son, Okinawa 904-0495, Japan\\
$^{2}$Dipartimento di Fisica and INFN, Universit\`a di Torino, 
via P. Giuria 1, 10125 Torino, Italy}

\begin{abstract}
We study the sedimentation of finite-size inertial particles in a
Rayleigh-Taylor-like setup using state-of-the-art direct numerical simulations.
The falling particles are observed to produce two distinct regions: a leading mixing layer with a linear concentration profile followed by a bulk region of uniform density. 
Unlike classical RT turbulence, the mixing layer extension accelerates with an anomalous, non-integer exponent, while the bulk region moves at a constant velocity.
A one-dimensional model based on a local hindered settling law accurately captures the observed
dynamics and its dependence on the particle-to-fluid density ratio. The present work
identifies a new regime of convective mixing 
which develops at the front of particle suspensions in sedimentation processes.


\end{abstract}


\maketitle 

{\it Introduction.} 
The problem of sedimentation has a long, rich history \cite{newton1687principia},
and its role in governing the natural world and shaping technical undertakings is well recognized \cite{lamb2020mud,koyaguchi1990sedimentation,meiburg-kneller2010,guazzelli2011physical}. Despite the deep-seated interest, however, the settling of particle collections remains an unsolved, formidable scientific challenge due to the dissipative nature of the fluid, the complex multi-body interactions and the emergence of a wide range of scales.

Among the most immediate, and yet still open, questions
is that of finding the average settling velocity $v$ of a suspension of solid particles. In fact, as particles settle, volume conservation produces an average upstream fluid motion.
This flow and its effects can be locally complex, and the overall sedimentation, when compared to the settling speed $v_t$ of a single particle, may be either hindered \cite{richardson1954sedimentation,batchelor1972,guazzelli2011physical} or enhanced \cite{huisman2016columnar}. For a suspension of solid fraction $\phi$, this problem is often referred to as that of finding the hindered settling function $f(\phi)=v/v_t$.

The initial distribution of the suspended particles (i.e. large scale spatial changes of $\phi$) can also affect the settling function.
One particularly relevant setup is that of a particle-laden suspension which sits atop the same clear, unladen fluid. 
This condition
is realized in a variety of natural
scenarios \cite{remington-arnett-ea1999,michioka2005instability,green1995plankton} and it is 
relevant in the context of riverine outflows, where a sediment rich river can
form a hypopycnal flow, i.e. a superficial layer of dispersed sediment overlaid
on a body of water \cite{wells-dorrell2021,parsons-bush-ea2001}. The unstable
particle-laden layer collapses, forming plumes of particles which feed
convective shear layers and eventually turbulence.  
In spite of the conceptual and practical relevance of the problem, fundamental
aspects remain scarcely investigated, with only few experimental
\cite{voltz-pesch-ea2001}, theoretical \cite{burns-meiburg2012} or numerical
\cite{chou-wu-ea2014,burns-meiburg2015,yu-hsu-ea2014,magnani-musacchio-ea2021}
works available mostly in the small inertia, high-dilution regime, where hindered settling effects are comparatively negligible. 

In this Letter, we investigate the yet unexplored regime of finite Stokes, finite dilution in the particle-laden Rayleigh-Taylor configuration, focusing our attention on the role of inertia. 
We use state-of-the-art, high-resolution direct numerical simulations to study the flow produced by the gravitational settling of $O(10^5)$ finite-size heavy particles layered above an otherwise quiescent fluid. 
We observe the formation of a turbulent mixing layer in the advancing interfacial
region with a quasi-linear vertical profile of particle concentration, followed
by a bulk region in which particle density remains constant in time. 
At variance with the classical RT phenomenology \cite{boffetta2017incompressible},
we find that the evolution of the extension of the mixing layer 
displays a power-law growth with an anomalous, non-integer exponent.
We develop a simple one-dimensional model of the hindered settling function
based on known physical ingredients, which is able to capture the
main features of the 
turbulent
sedimentation process. 

{\it Model and observables.}
We consider a large number $\ndp$ of monodisperse spherical particles settling in a still fluid in the presence of gravity ${\bf g}=(0,0,-g)$.
Particles have diameter $d_p$ and density $\rho_p$ larger than that of the fluid $\rho_f$. The density ratio is $\gamma=\rho_p/\rho_f$. 
At time $t=0$ the particles are randomly distributed exclusively in the top half of the domain, with an 
initial particle volume fraction of $\phi_b$.

The fluid phase is governed by the Navier-Stokes equations
\begin{equation}
{\partial {\bf u} \over \partial t} + {\bf u} \cdot {\bf \nabla u} =
{1 \over \rho_f} {\bf \nabla} p + \nu \nabla^2 {\bf u} + \frac{{\bf f}^{pf}}{\rho_f},
\label{eq1}
\end{equation}
for the incompressible velocity field ${\bf u}({\bf x},t)$. 
${\bf f}^{pf}$ represents the coupling with the particles, whose motion is given by the Newton-Euler equations of motion for rigid spheres
\begin{equation}
m_p {d {\bm v}_p \over d t}={\bf f}_g+{\bf f}^{fp}+{\bf f}^{pp},
\label{eq2}
\end{equation}
\begin{equation}
I_p {d {\bf \omega}_p \over d t}={\bf L}^{fp}.
\label{eq3}
\end{equation}
Here, $m_p=\rho_p V_p$ is the mass of the particle of volume $V_p$, $I_p=m_p
d_p^2/10$ its moment of inertia, ${\bm v}_p$ and ${\bf \omega}_p$ are the
particle's translational and rotational velocities, 
${\bf f}_g=(\gamma-1)\rho_f V_p {\bf g}$ the buoyancy force, 
${\bf f}^{fp}$ and ${\bf L}^{fp}$ are
respectively the force and torque due to the fluid-particle interaction,
obtained as the particle surface integrals of the coupling force density 
${\bf f}^{pf}$ and its moment with respect to the particle center 
${\bf r_p}\times {\bf f}^{pf}$.

Direct numerical simulations of the equations (\ref{eq1}--\ref{eq3}) are performed using the in-house developed finite-difference solver \emph{Fujin} (see Supplemental Material~\cite{supp}). The equations are discretised in a vertically elongated domain of size $L_x\times L_y\times L_z$, where $L_x=L_y$ and $L_z=4L_x$.
Periodic boundary conditions are imposed on the horizontal directions $(x,y)$, while no-slip, no-penetration conditions are set on the top and bottom of the domain. 
Following the system's symmetries, the collective dynamics of the particle ensemble can be described in terms of the vertical profiles of the local volume fraction $\phi(z,t)$ and of the
particle vertical velocity $v(z,t)$, averaged over the horizontal planes. The initial condition for our setup then corresponds to $\phi(z,0) = 0$ for $0 < z < L_z/2$ and $\phi(z,0) = \phi_b$ for $L_z/2 < z < L_z$, while $v(z,0)=0$ in all the domain. We set $\ndp=10^5$ corresponding to $\phi_b\approx0.10$.
We investigate four different values of the particle-to-fluid density ratio $\gamma={2,4,8,16}$. 
In order to develop a mathematical model for the mixing layer, we performed additional simulations at volume fractions $\phi_b=0.05$ and $\phi_b=0.15$.

\begin{figure}[h!]
\centerline{\includegraphics[width=\linewidth]{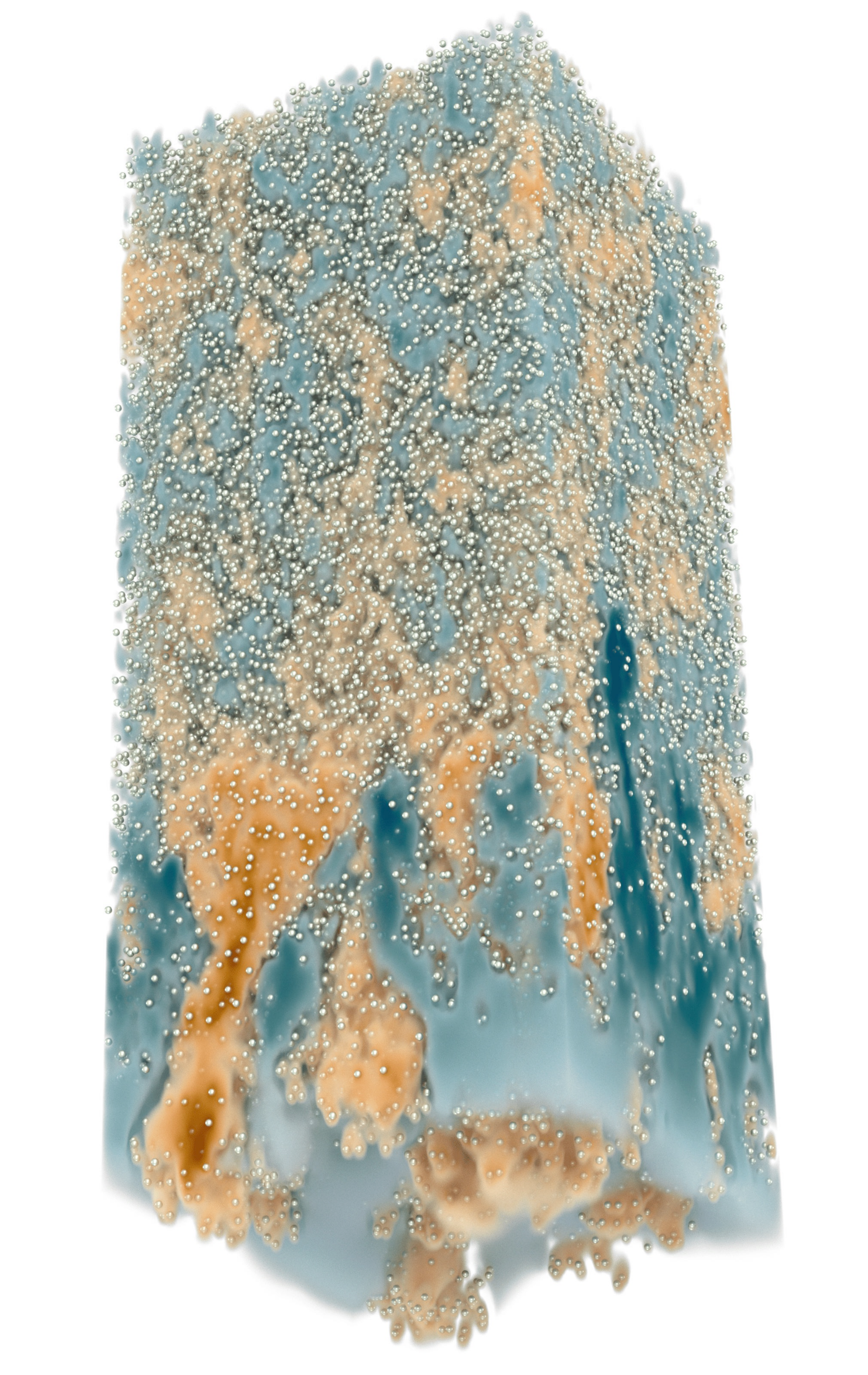}}
\caption{Volumetric rendering of the instantaneous vertical fluid velocity
field $v_z({\bm x},t)$ and particle positions at $t\approx0.17\tau_t$. 
Upward/downward velocities are shown in blue/orange, more intense
for higher magnitudes of $v_z$. Regions where $v_z\approx0$ are rendered as
transparent.}
\label{fig1}
\end{figure}

Figure~\ref{fig1} shows a snapshot of the system partway through its evolution for the $\gamma=2$ case. 
At time $t=0$ particles begin to fall from the upper part of the domain. The motion is however not uniform, and the formation of plume-like structures in the suspension causes particles closer to the interface with free fluid to fall faster. This produces a complex interfacial region in which the fluid becomes turbulent and affects the dynamics of the falling particles. 

\begin{figure}
\centerline{\includegraphics[width=0.95\linewidth]{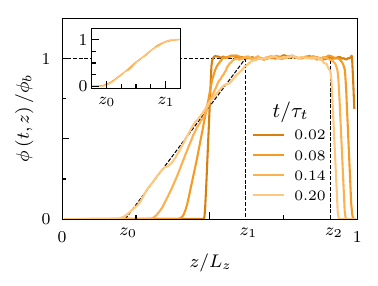}}
\caption{Particle concentration profiles for the $\gamma=2$ case at different
times. Dashed lines represent the profile model (\ref{eq4}) together with the positions of its characteristic points $z_0$, $z_1$ and $z_2$ for the latest time. In the inset, the latest three curves are rescaled according to the instantaneous locations of $z_0$ and $z_1$.}
\label{fig2}
\end{figure}

Figure~\ref{fig2} displays the vertical profile of particle concentration
$\phi(z,t)$ for the case $\gamma=2$ at different times.
In spite of the complexity of the dynamics, vertical profile has a relatively simple form with two distinct regions.
In the upper part $z_1(t) \le z \le z_2(t)$ (the ``bulk'') the volume fraction
remains uniform and equal to the original value $\phi_b$,  
while in the lower part
$z_0(t) \le z \le z_1(t)$ (the ``mixing region'') the particle-laden phase mixes with the unladen
fluid, resulting in a non-uniform concentration developing a close to linear 
profile. Both the bulk and the mixing regions fall but with 
different velocities, and therefore the shape of the profile changes in time. 
Based on these observations, we model the concentration profile with a simple
piecewise linear function
\begin{equation}
\phi(z,t) = \left\{
\begin{array}{ll}
 0 & z \le z_0 \land z > z_2 \\
\phi_b \frac{z-z_0}{z_1-z_0} & z_0 \le z \le z_1 \\
\phi_b & z_1 \le z \le z_2
\end{array}  
\right.
\label{eq4}
\end{equation}
From the points in (\ref{eq4}), we define the extension of the mixing 
region $h(t)=z_1-z_0$ and its central point 
$z_c(t)=(z_1+z_0)/2$. Volume conservation requires that the distance
$z_2-z_c$ remains constant.

\begin{figure}[h]
\centerline{\includegraphics[width=\linewidth]{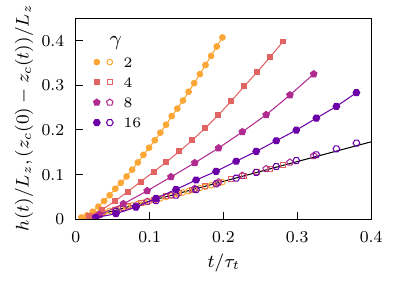}}
\caption{Time evolution of the width $h(t)$ (filled symbols) and geometric center $z_c(t)$ (empty symbols) of the mixing region for the simulations at different 
$\gamma$. For the geometric center, a black continuous line of slope $r$ is included for guidance.
}
\label{fig3}
\end{figure}

Figure~\ref{fig3} displays the time evolution of the width $h(t)$ and of the center $z_c(t)$ 
of the mixing region for the four cases at different density ratio.
For each run, the values of $z_0(t)$, $z_1(t)$ and $z_2(t)$ are obtained by a fit of the numerical data with the piecewise linear function (\ref{eq4}) at the different times. 
Figure~\ref{fig3} shows that while $z_c(t)$ (and therefore $z_2$) moves approximately
with a constant velocity $v_b$, proportional to the terminal velocity of a single particle $v_t$,
the growth of $h(t)$ is accelerated with respect to $v_t$. This effect is strongly dependent on the density ratio, and becomes more important for smaller values of $\gamma$.

The different velocities observed in the bulk and mixing regions
arise from differences in the suspension local structure.
In the bulk of the suspension, where particles remain homogeneously
distributed, the particle falling speed is reduced with respect to a single
particle because of the change in background flow speed due to the fluid rising
homogeneously through the bulk.  
By comparing the simulations at different volume fractions, we find that, 
this reduced hindered settling velocity is proportional to $\phi_b$, i.e.
$v_b \simeq v_t(a- d \phi_b)$ (where $a$ and $d$ are constants of order one independent
on $\gamma$, see Supplementary material), within the range of parameters explored. 
This is reminescent of the Richardson-Zaki (RZ) relation for the hindered 
settling velocity of Stokesian inertial particles which is valid for 
small particle Reynolds number $Re_p$ \cite{richardson1954sedimentation}.
Inside the mixing region, the particle velocity is not constant, as particles segregate into falling plumes (see Fig.~\ref{fig1}), and, as a result, are accelerated by wake effects. This is reflected in the $z$-dependence of the local volume fraction $\phi$. 

To account for these physical ingredients, we propose a simple 1D model for the
evolution of the relative concentration based on mass conservation and its
transport by the vertical velocity
\begin{equation}
{\partial \phi \over \partial t}=-{\partial \over \partial z} 
\left( v \phi \right),
\label{eq5}
\end{equation}
where $v$ is given by a 
local hindered model which generalizes the result observed 
for the bulk region 
\begin{equation}
v(z,t) =  - v_0 \left( 1 - c \phi(z,t) \right),
\label{eq6}  
\end{equation}
where $v_0$ is the velocity of the front position $x_0$ 
and $c=(v_0-v_b)/(v_0 \phi_b)$ represents 
the relative velocity difference between the front and the bulk of 
particle distribution. 

Assuming, as 
shown by the inset of Fig.~\ref{fig2}, that the concentration profile inside the mixing region is a self-similar function $\phi(z,t) = \phi_b f(\tilde{z})$ of the non-dimensional variable $\tilde{z} = (z-z_c(t))/h(t)$,  we obtain from (\ref{eq5}--\ref{eq6}) the relation 
\begin{equation}
\tilde{z} {dh \over dt} + {dz_c \over dt} = 
- \left[ v_0 + 2 (v_b-v_0) f(\tilde{z}) \right],
\label{eq7}  
\end{equation}
which shows that $f(\tilde{z})$ must be the linear function 
$f(\tilde{z}) = 1/2 + \tilde{z}$ in the mixing region 
$|\tilde{z}| \le 1/2$, which indeed corresponds to the profile (\ref{eq4}).

The center of the mixing region moves downward at the same speed of the bulk
$v_b$, while the growth of the mixing region is determined by the difference
between the velocity of the front $v_0$ and of the bulk $v_b$: 
\begin{equation}
\frac{d z_c}{dt} = -v_b \;, \;\;   \frac{d h}{dt} = 2 (v_0-v_b)
\label{eq8}
\end{equation}

The velocity difference $v_0-v_b$ is due to the turbulent motions inside the
mixing region, the intensity of which depends in general on $h$ and on the density 
ratio $\gamma$. By factorizing this dependence as 
$v_0-v_b \propto v_b (\gamma-1)^\beta (h/L_z)^\alpha$, and using the 
fact that $v_b=r v_t$, we finally obtain the predictions: 
\begin{equation}
\frac{z_c(t)-z_c(0)}{L_z} = -r \frac{t}{\tau_t} \;, \;\;
\frac{h(t)}{L_z} = b \left(\frac{t}{\tau_t}(\gamma-1)^\beta\right)^\xi ,
\label{eq9}
\end{equation}
where we have introduced the settling time of a single particle 
$\tau_t = L_z/v_t$, $\xi = 1/(1-\alpha)$, and $\beta$ and the integration constant $b$ (which depends on the initial volume fraction) are fitting parameters. 

\begin{figure}[h]
\centerline{\includegraphics[width=\linewidth]{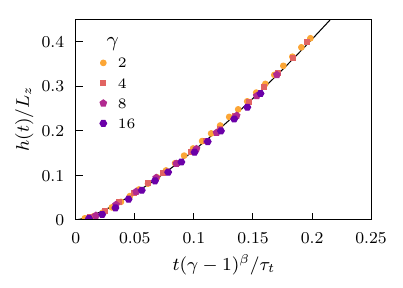}}
\caption{Time evolution of the width $h(t)$ of the mixing region.
Results from different $\gamma$ are rescaled according to the model (\ref{eq9}),
which is represented by the black continuous line.
}
\label{fig4}
\end{figure} 

We fit the time evolution of $h(t)$ shown in Fig.~\ref{fig3} 
with the prediction (\ref{eq9}). 
Remarkably, we find the values of the exponent $\xi$ to be independent of $\gamma$, and we compute its average value and the error as $\xi=1.375 \pm 0.015$. 
We find that all the runs are very well fitted with 
$\beta=-1/3$, which is therefore fixed, and the coefficient $b$ is 
obtained as $b = 3.75 \pm 0.10$. Incidentally, we note that the sustained acceleration of the mixing layer expansion ($\xi>1$) effectively translates into settling enhancement at late times, with $v_0$ growing up to $1.5v_t$ for case $\gamma=2$ within the investigated time span.

In Fig.~\ref{fig4} we plot the mixing layer width $h(t)$ rescaled according 
to the model using the above parameters 
together with the analytic prediction (\ref{eq9}). 
The remarkable quality of the rescaling indicates that our 1D model is indeed
able to capture the complex dynamics of the evolution of particle distribution in a
wide range of density ratio. 
The fact that the exponent $\beta$ of the density ratio $\gamma$ is
negative reflects quantitatively what observed in Fig.~\ref{fig3},
i.e., that the mixing layer is more accelerated with respect to the 
single particle velocity $v_t$ for lower density ratio.
Of course, this behavior cannot be extended to 
very small values of $\gamma-1$, because of the divergence of
(\ref{eq9}) as $\gamma \to 1$. Nonetheless, in the wide range of 
density ratio investigated by our simulations, the scaling exponent 
$\beta=-1/3$ is able to describe the observed phenomenology. 

The hindered model (\ref{eq5}--\ref{eq6}) allows us to make quantitative predictions for 
other important quantities, such as the particle flux. This is defined as 
\begin{equation}
\Pi(z,t)=-v(z,t) \phi(z,t),
\label{eq10}
\end{equation}
and represents the vertical flux of the solid phase. A reference value for this flux is that corresponding to a rigid motion of the bulk of particles with velocity $v_b$, that is $\Pi_b=v_b \phi_b$. The model prediction for the particle flux in the mixing region $|\tilde{z}|\le 1/2$ has the simple expression in the non-dimensional variable $\tilde{z}$
\begin{equation}
{\Pi(z,t) \over \Pi_b} = \left({1 \over 2}+\tilde{z} \right)+{v_0-v_b \over v_b} \left({1 \over 4}-\tilde{z}^2 \right),
\label{eq11}
\end{equation}
where we remark that both $\tilde{z}$ and $v_0$ are functions of time.

\begin{figure}[h]
\centerline{\includegraphics[width=\linewidth]{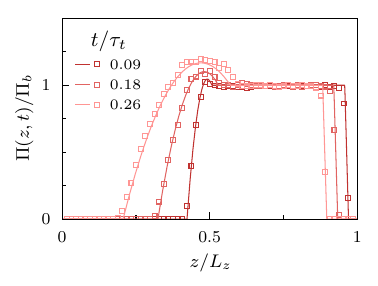}}
\caption{Vertical profile of the particle flux $\Pi(z,t)$ for the case $\gamma=4$ at different times. Lines represent the predictions of the theoretical model in the mixing layer given by (\ref{eq11}), markers represent the flux according to the simulation data.}
\label{fig5}
\end{figure} 

Figure~\ref{fig5} shows that the prediction (\ref{eq11}) reproduces well the numerical vertical flux obtained from our simulations. We remark that no additional parameter is fitted in (\ref{eq11}) since all the coefficients are fixed from the time evolution of $h(t)$.
A distinctive feature of the particle flux is the presence of a maximum just below the bulk region. This is due to particles accelerated from the lower boundary of the bulk region by the turbulent-like flow in the mixing region and is well captured by the model prediction (\ref{eq11}).


{\it Conclusions.}
We studied the dynamics of a layer of finite-size heavy particles 
sedimenting in an initially quiescent fluid. In spite of the 
complexity of the phenomenon, which produces a non-homogeneous 
turbulent-like flow, we find that, 
in the parameter space explored, 
the particle concentration profile 
follows a self-similar evolution with a linear mixing layer 
followed by a bulk region of constant concentration. 
Due to differences in the local structure of the suspension among the two regions, we observe both hindrance and enhancement of the average particle settling speeds.

Compared to the classic Rayleigh-Taylor phenomenology, which predicts quadratic scaling for $h(t)$, we find a superlinear but less-than-quadratic power law growth of the mixing layer, with scaling exponent  $\xi\approx1.37$ 
in the range of parameters investigated. 
To elucidate the effect of the observed dynamics, we introduce a one-dimensional transport model based on the observation of a RZ-like hindered settling relation within the bulk region, and its further local extension within the mixing layer, which we find justified \emph{a posteriori}. 
The model is able to explain both the anomalous scaling and its dependence on the density ratio, and is able to accurately predict quantities such as the vertical particle flux. 
Note that, the standard RT phenomenology is expected in the limit $d_p \to 0$ and $n_p \propto d_p^{-3}$, such that the dynamics of the particles recovers that of Lagrangian tracers and the density difference 
between the suspension and the pure fluid remains finite.  
This regime is outside the parameter space explored here. 


This work represents only a first step in understanding the dynamics of inertial suspensions at large concentrations. Future works will be required to better characterize the fundamental mechanics driving the anomalous mixing layer expansion. Moreover, there would be additional value in further extending our investigation to other regions of the parameter space, as well as comparing with the results of laboratory experiments.



\begin{acknowledgments}
\section*{Acknowledgments}
The research was supported by the Okinawa Institute of Science and Technology Graduate University (OIST) with subsidy funding to M.E.R. from the Cabinet Office, Government of Japan. M.E.R. also acknowledges funding from the Japan Society for the Promotion of Science (JSPS), grant 24K17210 and 24K00810. The authors acknowledge the computer time provided by the Scientific Computing \& Data Analysis section of the Core Facilities at OIST, and by HPCI, under the Research Project grant hp250021. S.T. acknowledges A. Chiarini and R. K. Singh for the helpful discussions while preparing this work.
\vspace{-0.5cm}
\end{acknowledgments}

%

\end{document}


\newcommand{\ndp}{n_{p}}
\newcommand{\ddp}{d_{p}}
\newcommand{\mdp}{m_{p}}
\newcommand{\Vdp}{V_{p}}
\newcommand{\Idp}{I_{p}}
\newcommand{\rhop}{\rho_{p}}
\newcommand{\rhof}{\rho_{f}}
\newcommand{\Gal}{\textit{Ga}}
\newcommand{\Atw}{\textit{A}}
\newcommand{\Sto}{\textit{St}}
\newcommand{\Rey}{Re}
\newcommand{\Ret}{\Rey_t}
\newcommand{\Rep}{\Rey_p}
\newcommand{\mlbot}{z_0}
\newcommand{\mltop}{z_1}
\newcommand{\mlcent}{c}
\newcommand{\bbot}{\mltop}
\newcommand{\btop}{z_2}
\newcommand{\bcent}{c_b}
\newcommand{\PhiV}{\Phi}
\newcommand{\Phitop}{\phi_{b}}
\newcommand{\taugrav}{\tau}
\newcommand{\taug}{\tau_g}
\newcommand{\taut}{\tau_t}
\newcommand{\vg}{v_g}
\newcommand{\vt}{v_t}
\newcommand{\mlwid}{h}
\newcommand{\bwid}{h_b}
\newcommand{\Zavg}{{xy}}
\newcommand{\ftop}{{fp}}
\newcommand{\ptof}{{pf}}
\newcommand{\ptop}{{pp}}
\newcommand{\gridsp}{\Delta x}
\title{Supplementary materials for: \\A new form of mixing in turbulent sedimentation}
\author{Simone Tandurella$^{1}$}

\author{Marco Edoardo Rosti$^{1}$}

\author{Stefano Musacchio$^{2}$}

\author{Guido Boffetta$^{2}$}

\affiliation{$^{1}$Complex Fluids and Flows Unit, Okinawa Institute of Science and Technology Graduate University, 1919-1 Tancha, Onna-son, Okinawa 904-0495, Japan
 \\
$^{2}$Dipartimento di Fisica and INFN, Universit\`a di Torino, 
via P. Giuria 1, 10125 Torino, Italy}

\begin{abstract}
\end{abstract}

\pacs{}

\maketitle 

\setcounter{table}{0}
\makeatletter 
\renewcommand{\thetable}{S\@arabic\c@table}
\makeatother

\setcounter{figure}{0}
\makeatletter 
\renewcommand{\thefigure}{S\@arabic\c@figure}
\makeatother

\setcounter{equation}{0}
\makeatletter 
\renewcommand{\theequation}{S\@arabic\c@equation}
\makeatother


\section{Domain and resolution}
For the cases presented in the main text, equations (1--3) are solved in a vertically elongated domain of size $L_x\times L_y\times L_z$, where $L_x=L_y=2\pi$ and $L_z=4L_x$. On the basis of the chosen spatial resolution, this corresponds to $384 \times 384 \times 1536$ collocation points.

Statistical convergence was improved through repeated realizations of the flow, and the results averaged. In particular, for each case of particle inertia investigated, three independent realizations of the flow were performed, differing in the initial uniformly random distribution of particles in the top half of the domain. 
For the purpose of the quantities investigated in this work, the cross-realization variance has been found to be sufficiently small as to be negligible. 


\subsection{Domain convergence}
To understand whether results are affected by confinement, we investigate the effect of increased domain size in both the homogeneous and non-homogeneous directions. We focus on the case $\gamma=16$, and test three different domain aspect ratios: normal ($N$), tall ($T$), and wide ($W$). The corresponding domain sizes are reported in Table~\ref{tab:domains}. The aspect ratio used for the results presented in the main body of the work is that of case $N$. Unlike in the main manuscript, the results of case $N$ correspond to a single realization of the flow.
In each case, at the outset the top half of the domain contains the particle-laden phase, while remaining flow parameters are fixed. This corresponds to a change in number of particles $\ndp$ across different domain sizes. 

\begin{table} [hb]
  \centering
  \caption{Domain sizes $L$, number of collocation points $N$ and number of particles $\ndp$ used for confinement tests.}
  \label{tab:domains}
  \begin{center}
    \begin{tabular}{c@{\hskip 0.2in}c@{\hskip 0.2in}c@{\hskip 0.2in}c}
      ID  & $L_x \times L_y \times L_z$ & $N_x \times N_y \times N_z$ & $\ndp$  \\
      \hline 
      N & $2\pi \times 2\pi \times\ \,8\pi $ & $384 \times 384 \times 1536$ & \num{100000} \\
      T & $2\pi \times 2\pi \times16\pi    $ & $384 \times 384 \times 3072$ & \num{200000} \\
      W & $4\pi \times 4\pi \times\ \,8\pi $ & $768 \times 768 \times 1536$ & \num{400000} \\
    \end{tabular}
\end{center}
\end{table}

In Figure~\ref{fig:dommixlayer} we showcase the time evolution of the mixing layer for the three different cases. To correctly compare the normalized width across different cases, the length scale chosen is that of $L_{z}$ of case $N$, that is, $L_z = 8\pi$. The growth of the mixing layer appears consistent across all the three cases. A deviation for long times is present for case $T$, though its magnitude appears relatively small and within the limits of cross-realization variance.

\subsection{Resolution convergence}
In order to gauge the accuracy of the chosen discretization in describing the system, we performed grid convergence tests. 
We focus on case $\gamma=2$, and test three different discretizations of the $2\pi \times 2\pi \times 8\pi$ domain. Based on the size of the grid, we call them small ($S$), medium ($M$), and large ($L$), which correspond to the resolutions reported in Table~\ref{tab:resolutions}. The discretization used for the results presented in the main body of the work is that of case $M$. The initial distribution of the particles is the same for all three resolutions. As in the case of domain convergence, also in this one, we compare only one realization of the three cases.

\begin{table} [hb]
  \centering
  \caption{Resolutions of simulations used for convergence test. We report the domain size $L$, the number of collocation points $N$ and the ratio of particle size to grid spacing $\ddp/\Delta x$.}
  \label{tab:resolutions}
  \begin{center}
    \begin{tabular}{c@{\hskip 0.2in}c@{\hskip 0.2in}c@{\hskip 0.2in}c}
      ID  & $L_x \times L_y \times L_z$ & $N_x \times N_y \times N_z$  &  $\ddp/\Delta x$   \\
      \hline 
      S & $2\pi \times 2\pi \times 8\pi $& $256 \times 256 \times 1024$ & 4.0 \\
      M & $2\pi \times 2\pi \times 8\pi $& $384 \times 384 \times 1536$ & 6.0 \\
      L & $2\pi \times 2\pi \times 8\pi $& $512 \times 512 \times 2048$ & 8.0 \\
    \end{tabular}
\end{center}
\end{table}

\begin{figure}
\centering
\begin{subcaptiongroup}
\subcaptionlistentry{}
\label{fig:dommixlayer}
\begin{overpic}[width=0.495\textwidth]{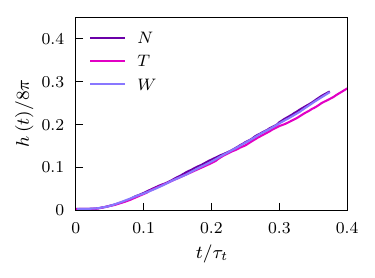}
\put(05,70){\captiontext*{}}
\end{overpic}%
\hfill
\subcaptionlistentry{}
\label{fig:resmixlayer}
\begin{overpic}[width=0.495\textwidth]{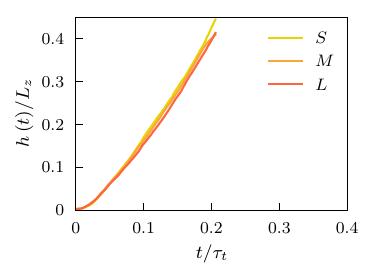}
\put(05,70){\captiontext*{}}
\end{overpic}
\end{subcaptiongroup}
\captionsetup{subrefformat=parens}
\caption{ Time history of the normalized mixing layer width for the \subref{fig:dommixlayer} domain convergence cases and \subref{fig:resmixlayer} resolution convergence cases.}
\end{figure}

In Figure~\ref{fig:resmixlayer}, we report the results of grid convergence for the mixing layer width.  
All cases show the same qualitative trend and very close magnitudes. Across the three realizations, we observe a possible tendency towards smaller width with higher resolution. However, its magnitude appears to be sufficiently small as to be comparable with statistical fluctuations. 

\section{Non-dimensional parameters}
The four different ratios of particle to fluid density $\gamma$ investigated can be related to other parameters that are commonly used in the domains of Rayleigh-Taylor and settling studies. Having set the fluid kinematic viscosity constant to $\nu=0.002$, the obtained $\Gal$, $\Atw$ and $\Ret$ are defined and reported in Table~\ref{tab:adim}.

\begin{table}[!h]
  \centering
  \caption{Corresponding values of non-dimensional quantities for the values of $\gamma$ investigated.
    $\Atw=\left(\rho_2-\rho_1\right)/\left(\rho_2+\rho_1\right)$, where $\rho_2=\rhop\Phitop+\rhof\left(1-\Phitop\right)$ and $\rho_1=\rhof$. $\Gal=\sqrt{(\gamma-1)\ddp g}\ddp/\nu$, and $\Ret=\vt\ddp/\nu$}
  \label{tab:adim}
  \vspace{.3cm}
  \begin{center}
    \begin{tabular}{c@{\hskip 0.2in}S@{\hskip 0.2in}S@{\hskip 0.2in}S@{\hskip 0.2in}S}
      $\gamma$      &  2.0 &  4.0 &  8.0 & 16.0 \\
      \hline
      $\Atw$          & 0.048& 0.130& 0.259& 0.429 \\
      $\Gal$          & 15.4 & 26.7 & 40.7 & 59.6 \\
      $\Ret$        & 7.94 & 18.1 & 33.3 & 56.9 \\
    \end{tabular}
\end{center}
\end{table}
The results are made dimensionless using the vertical scale $L_z$ as unit of length. We use particle terminal velocity of a single particle falling in an otherwise
unperturbed flow to make velocity dimensionless. Its value is obtained from an additional set of numerical simulations with the same parameters of the main runs but with a single particle starting from the top of the domain and the asymptotic
vertical velocity $v_t(\gamma)$ is measured after a short transient.
Finally, time unit is given by $L_z/v_t$.

\section{Volume fraction variation}

\begin{figure}
\centering
\begin{subcaptiongroup}
\subcaptionlistentry{}
\label{fig:phib_mixlayer}
\begin{overpic}[width=0.495\textwidth]{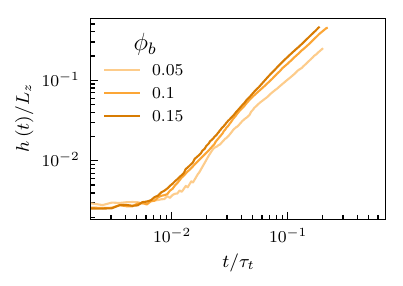}
\put(05,70){\captiontext*{}}
\end{overpic}%
\hfill
\subcaptionlistentry{}
\label{fig:phib_center}
\begin{overpic}[width=0.495\textwidth]{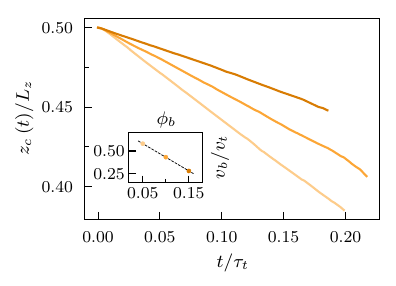}
\put(05,70){\captiontext*{}}
\end{overpic}
\end{subcaptiongroup}
\captionsetup{subrefformat=parens}
\caption{Effect of the variation of initial volume fraction $\phi_b$ for the case $\gamma=2$ on \subref{fig:phib_mixlayer} the mixing layer width $h$, and \subref{fig:phib_center} the location of the mixing layer centre $z_c$. In the inset of \subref{fig:phib_center}, we show the dependency of the bulk velocity $v_b$ on $\phi_b$, together with the line of best fit.}
\end{figure}

In order to test the robustness of the modeling, we performed additional simulations at different initial volume fraction $\phi_b$ for the case $\gamma=2$. In particular, we set $\phi_b=0.05$ and $\phi_b=0.15$. In Figure~\ref{fig:phib_mixlayer} we plot the mixing layer growth for the three cases. We find that the late-time temporal scaling exponent $\xi$ does not appear to appreciably change with variation of $\phi_b$. Additionally, in Figure~\ref{fig:phib_center} we show the position of the center for the different cases. We find the rate of descent $v_b=d z_c/dt$ to vary linearly with $\phi_b$ as $v_b=v_t(a-d \phi_b)$ with $a \simeq 0.8$ and $d \simeq 3$. A separate assessment of the bulk settling speed confirms that the rate of descent of the mixing layer center remains equal to the settling bulk velocity $v_b$ for all values of $\phi_b$.

\section{Numerical method}

The simulations presented in the work are performed using \emph{Fujin}, a in-house finite-difference incompressible fluid solver which features a second-order central finite difference scheme in space and second order Adams-Bashforth in time. The incompressibility constraint is enforced through a fractional step method.

Fluid-particle coupling is obtained through the use of the Eulerian Immersed Boundary Method (IBM) developed by \citet{hori-rosti-ea2022}. The particle position is tracked as a single Lagrangian point, while its footprint on the color function is discretized directly on the fluid simulation grid through the use of a fast hyperbolic tangent digitizer. The method includes implicit lubrication and an additional soft particle collision model. It has been extensively validated (\url{https://www.oist.jp/research/research-units/cffu/validation}) and used in previous works on particle-laden flows \citep{chiarini-cannon-ea2024,chiarini-tandurella-ea2025}. 
Despite the presence of a collisional contacts model, we find that, in the range of parameters investigated in this Letter, collisions do not play any meaningful role; indeed, at volume fraction $10\,\%$ and $\gamma=2$ we observe that only a fraction of particles smaller than $0.001$ is in contact at any instant. This fraction grows at most to $\approx0.018$ for $\gamma=16$. 


\section{Fitting the concentration profiles}

The vertical profile of the volume fraction $\phi(z,t)$ is defined as $\phi(z,t)\equiv\langle \Phi({\bm x},t) \rangle_{xy}$, where $\Phi({\bm x},t)$ is the particle field color function (equal to 1 inside the particles and 0 outside), and $\langle \cdot \rangle_{xy}$ represents the average over the horizontal planes of the domain. A similar definition is used for the particle vertical velocity $v(z,t) \equiv \langle \Phi({\bm x},t) v_z({\bm x},t)\rangle_{xy} / \phi(z,t)$. 

The numerical concentration profiles $\phi(z,t)$ are fitted, for each time, by the piecewise linear function
\begin{equation}
\phi(z,t) = \left\{
\begin{array}{ll}
 0 & z \le z_0 \land z > z_2 \\
\phi_b \frac{z-z_0}{z_1-z_0} & z_0 \le z \le z_1 \\
\phi_b & z_1 \le z \le z_2 \\
\end{array}  
\right.
\label{eq1}
\end{equation}
with all the parameters $\phi_b(t)$ and $z_i(t)$ free at each time.  
We find that value of $\phi_b\approx 0.103$ to be independent on time, showing no compaction nor expansion of the bulk during the evolution of the system. 

Due to the discrete nature of the particles, the measured spatial profiles of volume fraction and velocity present small scale noise. In order to correctly represent the salient features of the data, we smooth out the small scale noise through a coarse-graining procedure.  
In particular, we apply a fixed \num{33} grid-point (corresponding to $\num{0.02}L_z$ or $\num{5.5}\ddp$) rolling average stencil to the measured $\phi(z,t)$ and $\Pi(z,t)$ presented respectively in Fig.~2 and Fig.~5 of the main text. 
We remark that this coarse-graining procedure is performed only for display purposes, while the raw data is used for the fitting procedure.
\bibliography{biblio}